# Ferrimagnetism in stable non-metal covalent organic framework

Dongge Ma[1,*], Yuhang Qian[1], Mingyang Ji[1], Jiani Li[1], Jundan Li[1], Anan Liu[3], Yaohui Zhu[2,*]

[1]College of Chemistry and Materials Engineering, Beijing Technology and Business University, Beijing, China

[2]Physics Department, School of Artificial Intelligence, Beijing Technology and Business University, Beijing, China

[3]Basic Experimental Centre for Natural Science, University of Science and Technology Beijing, Beijing, China

*Corresponding author: madongge@btbu.edu.cn, yaohuizhu@gmail.com

**Abstract:** We synthesized a pure organic non-metal crystalline covalent organic framework TAPA-BTD-COF by bottom-up Schiff base chemical reaction. And this imine-based COF is stable in aerobic condition and room-temperature. We discovered that this TAPA-BTD-COF exhibited strong magneticity in 300 K generating magnetic hysteresis loop in M-H characterization and giant $\chi_{mol}$ up to 0.028. And we further conducted zero-field cooling and field-cooling measurement of M-T curves. The as-synthesized materials showed a large $\chi_{mol}$ up to 0.028 in 300 K and increasing to 0.037 in 4.0 K with 200 Oe measurement field. The TAPA-BTD-COF $1/\chi_{mol} \sim T$ curve supported its ferrimagnetism, with an intrinsic Δ temperature as -33.03 K by extrapolating the $1/\chi_{mol} \sim T$ curve. From the continuously increasing slope of $1/\chi_{mol} \sim T$, we consider that this TAPA-BTD-COF belongs to ferrimagnetic other than antiferromagnetic materials. And the large $\chi_{mol}$ value 0.028 at 300 K and 0.037 at 4.0 K also supported this, since common antiferromagnetic materials possess $\chi_{mol}$ in the range of $10^{-5}$ to $10^{-3}$ as weak magnetics other than strong magnetic materials such as ferrimagnetics and ferromagnetics. Since this material is purely non-metal organic polymer, the possibility of d-block and f-block metal with unpaired-electron induced magnetism can be excluded. Besides, since the COF does not involve free-radical monomer in the processes of synthesis, we can also exclude the origin of free-radical induced magnetism. According to recent emerging flat-band strong correlated exotic electron property, this unconventional phenomenon may relate to n-type doping on the flat-band locating in the CBM (conduction band minimum), thus generating highly-localized electron with infinite effective mass and exhibiting strong correlation, which accounts for this non-trivial strong and stable ferrimagneticity at room-temperature and aerobic atmospheric conditions.

### I. Introduction

Recently, as the emerging of twisted bilayer magic-angle graphene (TWBMG)(*1, 2*), the exotic physics phenomena associated with flat-band induced strong electron correlation(*3*), such as superconductivity(*4, 5*), ferromagnetism(*6*), topological states(*7*) and spin Hall effect(*8*) have been investigated in low temperature(*9*) (mK level) and high magnetic field. Considering the difficulty to obtain a low-temperature condition to mK, it is important to level up the $T_c$ of such flat-band materials. Inorganic materials such as TWBMG(*10*) are emerging as potential flat-band materials. However, due to their narrow band-gap, and the considerable challenges to precisely tune and control the twisting angle to several degrees, it is increasingly important to develop tunable organic flat-band materials with exotic electronic property such as magnetism.

As a kind of novel emerging star materials, covalent organic framework (COF) is a type of pure organic polymeric material connected by strong covalent bond in two or three-dimension other than metal-organic framework (MOF), which is connected by weak coordinative bond(*11*). Besides, COF is crystalline polymer similar as MOF, mostly with crystalline domain as large as several hundred nanometers to several micrometers, even in certain structures can be prepared into 50-100 μm large single crystal(*12*). Apart from these, because of its covalent bond connection, COFs possess much higher chemical stability other than MOFs in aerobic humid, aqueous and organic solvents condition(*13*). Moreover, due to its crystallinity, the COFs structure can be certainly confirmed via the assignment of XRD data with Pawley or Rietveld refinement accompanying theoretical modelling and calculation(*14*). Their clear and unique structure facilitates the demonstration of QSAR (Quantitative structure–activity relationship) and guides the design and synthesis of the materials with the optimal property as desired(*15*). Since COFs are synthesized via bottom-up organic synthetic routes from organic small molecule monomers, the main-chain, branch-chain and pore structures can all be precisely tuned by choosing different monomers with diversified electronic and steric configurations. Besides, different dynamic chemical reactions such as boronate(*16*), Schiff-base(*17*), ketoenamine(*18*) and C=C(*19, 20*) formation reactions can be chosen to optimize COFs with best chemical, physical or biological application performances.

Since in 2005, Yaghi et al. discovered the first COF materials COF-1 and COF-5(*16*), COFs materials have been utilized in various fields and applications, such as gas adsorption/separation(*21*), catalysis(*22*), energy storage(*23*), proton conduction(*24*), fluorescence sensing(*25*) and optoelectronics(*26*). However, using COFs in magnetism as magnetic ordered materials is still rare. Jiang et al. reported in 2017 that an iodine-doped C=C linked sp$^2$c-COF can generate $\chi_{mol}$ up to 0.025 in 8 K(*19*). The authors demonstrated that it is a paramagnetic material other than ferromagnetic or ferrimagnetic because of the absence of a M-H hysterias loop and spontaneous magnetization phenomena, and this material exhibited paramagnetic property with measurable $T_c$ curie temperature. Wang(*27*), Cui(*28*), Liu(*29*) and Brédas(*30*) independently predicted the potential magnetism in Lieb-lattice by theoretical modelling and calculation. However, the macroscopic experimental realization of strong magnetism such as ferrimagnetism in pure organic non-metal materials is still not achieved. Herein, we demonstrated that a novel TAPA-BTD-COF, which is prepared via a Schiff-base condensation, exhibits large $\chi_{mol}$ up to 0.028 in 300 K and 0.037 in 4.0 K in aerobic atmospheric condition with 200 Oe measurement field. Moreover, this COF exhibits measurably considerable large hysterias loop in M-H curve in 300 K. By extrapolating and analyzing the $1/\chi_{mol} \sim T$ curve, we conclude that the TAPA-BTD-COF is a ferrimagnetics material with $\chi_{mol}$ ranging from 0.028 to 0.037 from 300 K cooling down to 4.0 K with 200 Oe measurement field. This is the first room-temperature strong magnetic organic materials exhibiting ferrimagnetism with giant $\chi_{mol}$ = 0.028 at 300 K in aerobic atmospheric condition which also possesses M-H hysteresis loop in room-temperature and large $\chi_{mol}$ = 0.037 in 4.0 K, while previous model organometallic magnetic material $V(TCNE)_x\gamma(CH_2Cl_2)$ possesses a $\chi_{mol}$ equal to $8*10^{-6}$ in 4.2 K and decompose in aerobic condition with short lifetime(*31*).

## II. Experimental and Methods Details

TAPA-BTD-COF was prepared with a mixture of tris(4-aminophenyl)amine (TAPA) (29.0 mg, 0.10 mmol), 4,4-(benzothiadiazole-4,7-diyl)dibenzaldehyde (BTD) (51.4 mg, 0.15mmol), n-butanol (2.0 mL), and 1,4-dioxane (1.0 mL), and acetonitrile (0.05 mL) charging in a cylindrical glass tube (20 cm of length, $\phi_{in}$ = 0.8 cm, $\phi_{out}$ = 1.0 cm) and sonicated for 30 min to get a homogeneous dispersion solution. Then 0.1 mL of 2 M aqueous acetic acid was dropwise added to the solution, along with the color change to dark red. After it was degassed by the typical three freeze−pump−thaw cycles with liquid nitrogen, the tube was sealed and then heated at 60 °C for 3 days. The dark red precipitate was collected by Soxhlet extraction and washed continuously with acetone and methanol. Finally, solids were washed and immersed in acetone for 24 h. After filtration, samples were dried in vacuum oven under 120 °C overnight. Yield: 90 %.

Powder X-ray diffraction data were collected using a Panalytical Empyrean diffractometer in parallel beam geometry employing Cu K$_\alpha$ line focused radiation ($\lambda$ =1.5405 Å) at 1600 W (40 kV, 40 mA) power. Sample powders were placing on glass substrate recorded from $2\theta$ =1.5° up to 30° with 0.02° increment.

Solid-state nuclear magnetic resonance (SS-NMR) spectra were collected on a Bruker AVANCE III 400 NMR spectrometer using a standard Bruker magic angle-spinning (MAS) probe with 4-mm zirconia rotors. The magic angle was adjusted by maximizing the number and amplitudes of the signals of the rotational echoes observed in the $^{79}$Br MASFID signal from KBr. The transmitter frequency of $^{13}$C NMR is 100.39MHz. The solid-state $^{13}$C NMR spectra were acquired using cross-polarization (CP) MAS technique with the ninety-degree pulse of $^1$H with 4μs pulse width. The CP contact time was 3 ms. High power two-pulse phase modulation (TPPM-15) $^1$H decoupling was applied during data acquisition. The decoupling frequency corresponded to 62.5 kHz. The MAS sample spinning rates were 8 kHz. Recycle delays between scans were varied from 2 to 2.5 s. The $^{13}$C chemical shifts are given relative to neat tetramethylsilane as zero ppm, calibrated using the methylene carbon signal of adamantane assigned to 38.48 ppm as secondary reference.

Fourier-transform infrared (FTIR) spectra of starting monomer materials and COF samples were recorded from 400 to 4000 cm$^{-1}$ by using KBr pellets on a Bruker Tensor-27 Fourier-transform infrared spectrometer

Solid-state UV-Vis electron absorption spectra of starting materials monomers and COF sample were recorded on a Perkin-Elmer Lamda diffuse-reflectance-spectrometer.

Scanning electron microscopy and its mapping-mode characterization of COF samples were measured by dispersing the materials onto silica wafers attached to a flat aluminum sample holder, which were further coated with platinum. Samples were analyzed on a Zeiss Sigma 300 field-emission SEM operating at 10 kV.

Elemental analyses were carried out on a Thermo Flash Smart CHNOS organic element analyzer.

Electron-paramagnetic-resonance (EPR) measurements were carried out using an X-band spectrometer, Bruker E500 in 300 K under aerobic atmospheric conditions.

Magnetization measurements were performed using a commercial superconducting quantum interference device (SQUID) magnetometer (MPMS-3, Quantum Design). All the magnetization data of TAPA-BTD-COF powder crystals were corrected by subtracting diamagnetic susceptibility of empty sample holder in atmospheric air condition.

Modeling of crystal structures for target COFs were performed using the Vesta (Visualization and Electronic Structural Analysis) Version 3.5.2 software (K. Momma and F. Izumi, "VESTA 3 for three-dimensional visualization of crystal, volumetric and morphology data," J. Appl. Crystallogr., 1272-1276 (2011)(*32*)). DFT calculations were carried out within the framework of the Perdew–Burke–Ernzerhof generalized gradient approximation (PBE-GGA), as embedded in the Vienna ab initio simulation package code(*33*). All the calculations were performed with a plane-wave cutoff energy of 500 eV. For the TAPA-BTD-COF, we adopted the experimental lattice constants and transferred to the primitive cell with a= 45.05700 Å; b= 45.05700 Å; c=3.65000 Å, α= β= 90°, γ=120° [full structural relaxation would only change the lattice constants negligibly <0.1%)]. TAPA-BTD-COF were optimized to have AA stacking which will be used for our study of both bulk and bilayer systems. The geometric optimizations were performed without any constraint until the force on each atom is <0.01 eV·Å$^{-1}$ and the change of total energy is smaller than $10^{-4}$ eV per unit cell. The Γ centered Brillouin zone k-point sampling was set with a spacing of 0.03 × 2π·Å$^{-1}$, which corresponds to 3 × 3 × 11 k-point meshes for bulk unit cell.

### III. Results and discussions

We condensed the two monomers tris(4-aminophenyl)amine (TAPA) and 4,4'-(benzo(1,2,5)thiadiazole-4,7-diyl)dibenzaldehyde (BTD) to form imine-connected TAPA-BTD-COF via a Schiff-base reaction. To improve its crystallinity and the reaction yield, we optimized the preparation conditions by screening multiple parameters. After tedious exploration, the optimum preparation conditions were identified, and we obtained the dark red crystalline powder with 90% yield (the synthetic route and structure details are shown in Fig. 1). The successful preparation of the proposed COF was confirmed by a series of characterization techniques including powder X-ray diffraction (PXRD), solid-state-cross-polarization magic-angle-spinning $^{13}$C-NMR (SS-CP-MAS-$^{13}$C-NMR), Fourier-transform infrared spectrometry (FT-IR), element analysis (EA), scanning electron microscopy (SEM) and element-mapping (See Fig. 2 and 3). The COF's structure was first characterized by PXRD combining with theoretical modelling and simulations (see Fig 2A). TAPA-BTD-COF exhibited six diffraction peaks in PXRD spectrum, at $2\theta$ = 2.20º, 3.86º, 4.46º, 5.93º, 7.82º and 24.73º. With the proposed structure and the PXRD pattern at hand, we conducted the theoretical calculation to simulate the minimum-energy geometry configuration via the DFT methods by modelling with VESTA 64 programs and calculating with VASP software package (see experimental and methods details part). The theoretical calculations indicated that the AA-stacking structure possessing the following diffraction peaks at 2.26º, 3.92º, 4.53º, 5.99º, 7.84º and 24.36º well matched the experimental spectrum. Then, the full-profile Rietveld refinement using GSAS software was conducted. As shown in Fig.2A, the refined AA-stacking structure provided satisfactory matching results between calculated curves and refined experimental spectra with small $R_{wp}$=2.47% and $R_p$=1.80%. From the experimental crystallographic results, the TAPA-BTD-COF lattice parameters were determined as a= 46.70 Å, b= 45.06 Å, c= 3.65 Å, $\alpha=\beta$= 90.000°, $\gamma$= 120.000°. Furthermore, the COF chemical structure was investigated by SS-CP-MAS-$^{13}$C-NMR. There appear four prominent $^{13}$C signals peaking at 153.78, 145.26, 136.74 and 129.37 ppm of TAPA-BTD-COF (see Figure 2B). Based on the previous reporte TAPA and BTD-containing COFs structures, the corresponding carbons in different chemical environments were assigned. The peak at the most downfield (153.78 ppm) is belonged to the imine C=N and benzothiadiazole C=N carbons (marked as a and b in Figure 2B inbox). The peak at 145.26 ppm originates from the quaternary carbons in the TAPA and phenyl rings. The peak centered at 136.74 ppm corresponds to the quaternary and tertiary carbons of the phenyl rings between TAPA and BTD rings. The peak at the upper-most field (129.37 ppm) corresponded to the ternary carbons in the TAPA rings, benzothiadiazole carbons and the phenyl g-position carbons signals. Moreover, thorough disappearance of signals at 196.77 ppm of C=O indicates that the aldehyde is totally consumed up. With the new emerging signal at 153.78 ppm of imine C=N carbon together, the formation of the imine-linked TAPA-BTD-COF was further confirmed. The FTIR spectrometry was conducted to characterize the formation of the imine functional group and the consummation of the amine and carbonyl functional groups. As shown in Fig. 2C, after condensation reaction, the TAPA amine N-H stretching vibration signals at 3336 and 3406 $cm^{-1}$ wavenumbers disappeared. Meanwhile, both the aldehyde H-C=O stretching vibration at 2720 and 2820 $cm^{-1}$ and C=O stretching vibration at 1687 $cm^{-1}$ wavelength of BTD precursor disappear in newly formed COF FTIR spectrum. Besides, the C=N imine stretching vibration at 1617 $cm^{-1}$ appeared in the COF spectrum whereas both TAPA and BTD precursors spectra lacked this signal. These results further evidenced the formation of imine COF from TAPA and BTD. The electronic absorption spectra of the three COFs studied in this work were characterized via diffuse reflectance spectrometry. As shown in Fig. 2D, the obtained TAPA-BTD-COF exhibited red-shift in their absorption in comparison with the TAPA and BTD monomers. This result demonstrated that the formation of imine-based COFs structures brought about the higher degree of conjugation, which widened their UV-Vis absorption range. To obtain the information of their microscopic morphology, the SEM and element mapping were undertaken (see Fig 3). From the SEM images, we observed that the TAPA-BTD-COF exhibited ordered spherical nanoparticle morphology with 500-800 nm diameter as shown in Fig 3A. The element-mapping images by SEM demonstrates that the C, N and S elements distributed homogeneously in the whole COF nanocrystals (see Figure 3B-F). Furthermore, the chemical compositions of the as-synthesized TAPA-BTD-COF, was confirmed by the element analysis (EA) technique. The obtained EA experimental results well agreed with the theoretical calculated value based on the proposed COF structures (see Table 1). The experimental weight composition of TAPA-BTD-COF is C (75.07%), H (4.39%), N (12.19%), and S (5.78%), while the infinite two-dimensional layer model is (calculated for a single cell: $C_{96}H_{60}N_{14}S_3$) C 76.59%, H 3.99%, N 13.03%, and S 6.38%).

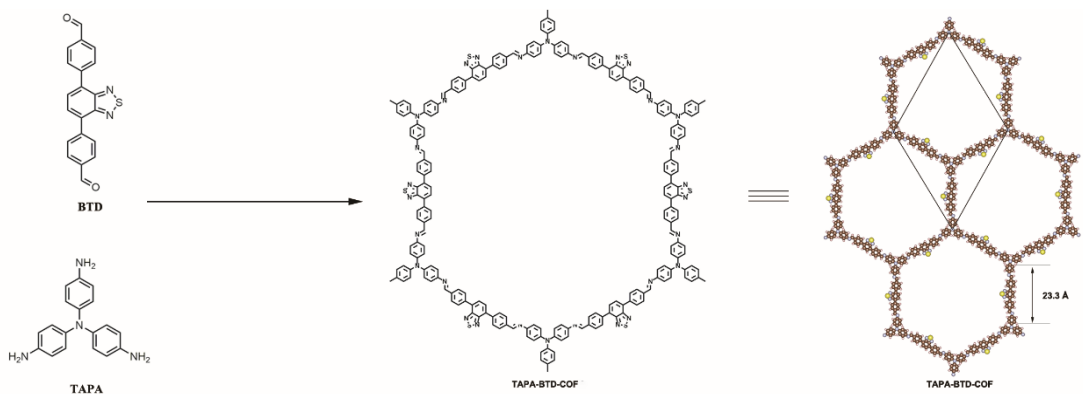

**Figure 1.** The schematic representation of the chemical structures of TAPA-BTD-COF

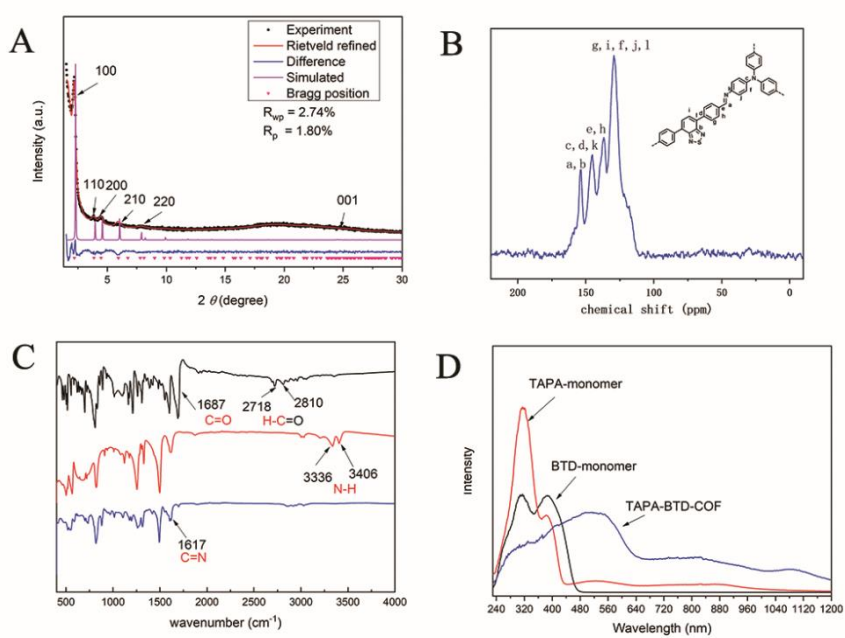

**Figure 2.** Structural characterization of TAPA-BTD-COF (A) X-ray-diffraction (XRD) analysis of TAPA-BTD-COF with the experimentally observed pattern in black, the Rietveld full-profile refined curves in red, and the difference plotted in blue. (B) Solid-state $^{13}$C CP-MAS NMR spectrum of TAPA-BTD-COF. (C) Fourier transform-infrared spectrum of (C) TAPA-BTD-COF, blue; TAPA monomer, red; BTD monomer, black. (D) Solid-state UV-Vis electron-absorption spectrum of TAPA-BTD-COF, blue; TAPA monomer red; BTD monomer, black;

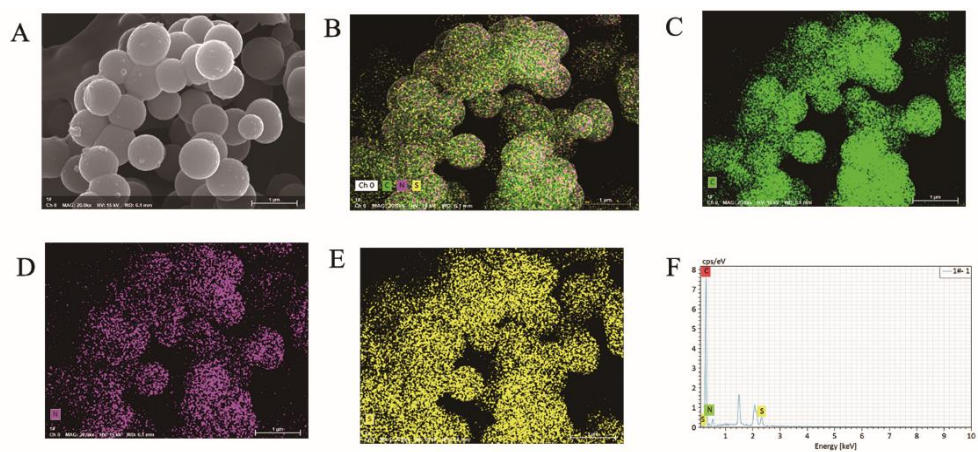

**Figure 3.** Scanning electron microscopy of TAPA-BTD-COF (A) SEM image of TAPA-BTD-COF (B) C, N and S element total mapping in SEM-mapping mode. (C) Separate C element distribution in SEM-mapping. (D) S element distribution in SEM-mapping. (E) S element distribution in SEM-mapping. (F) The energy spectrum in SEM-mapping.

**Table 1.** Element analysis results of TAPA-BTD-COF

| | | | | |
|---|---|---|---|---|
| Theoretical value | C (76.59%) | H (3.99%), | N (13.03%) | S (6.38%) |
| Experimental value | C (75.07%) | H (4.39%) | N (12.19%) | S (5.78%) |

From the deficiency of sulfur element as shown in element analysis results between experimentally obtained TAPA-BTD-COF material and theoretically modelled perfect TAPA-BTD-COF crystal structure (see Table 1), we considered that this S deficiency also represents the deficiency in BTD unit compared with TAPA unit. As well known, sulfur-containing BTD (benzothiadiazole) unit is extremely electron-poor, while TAPA (tris-4-aminophenyl-amine) is an electron-rich group. The deficiency in BTD indicates that the COF possesses excess electrons compared with its intrinsic state.

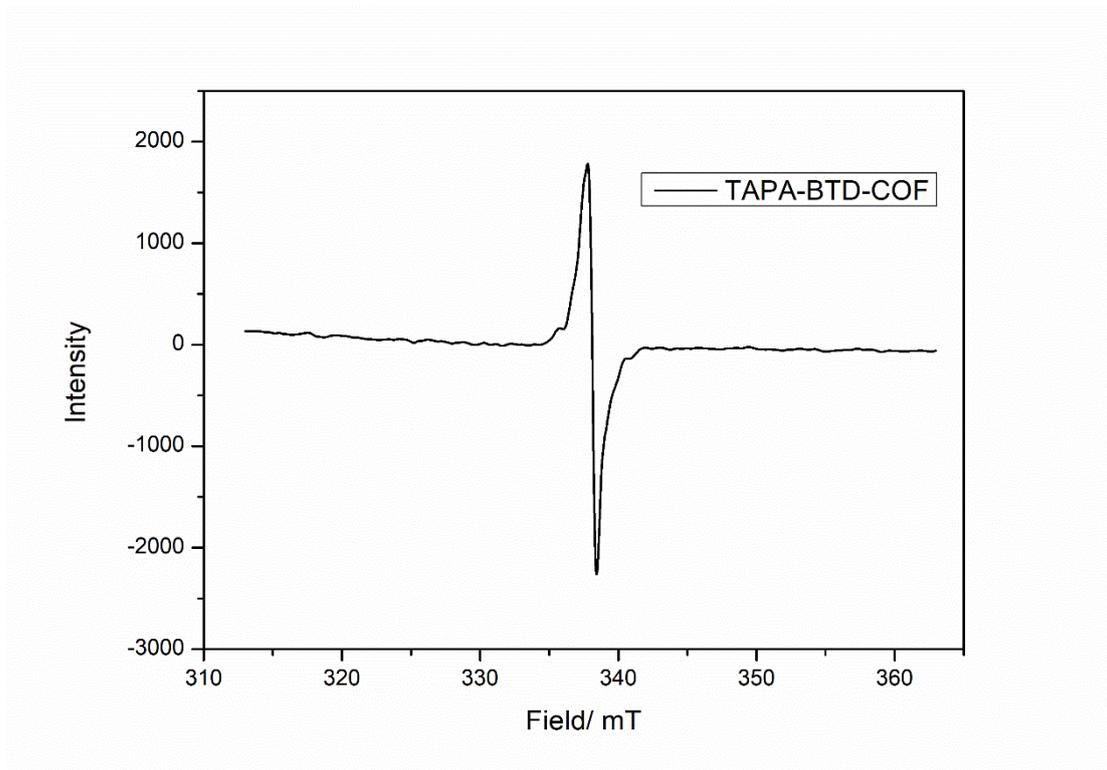

**Figure 4.** Electron-paramagnetic-resonance (ESR) spectrum of TAPA-BTD-COF in 300 K under aerobic atmospheric conditions.

Based on such consideration, we conducted the band-structure calculation and analysis. To our delight, in the intrinsic TAPA-BTD-COF band structure, there appear a set of 3 bands in including the conduction band bottom (CBM) just above the Fermi level. The three bands include two Dirac bands with a crossing point and a flat band (conduction band) under these two Dirac bands (see Figure 5). The VBM bands contain Right below the Fermi level is a set of two Dirac bands with a crossing point (Dirac cone). The appearance of non-dispersive flat band means the existence of infinite heavy electron with zero-velocity. These largely localized zero-velocity electrons is evidenced to belong to strong-correlated electron system. The strong correlation between electrons in flat band would generate various exotic non-trivial phenomena such as ferromagneticity, ferrimagneticity, antiferromagneticity, superconductivity, topological states, etc.

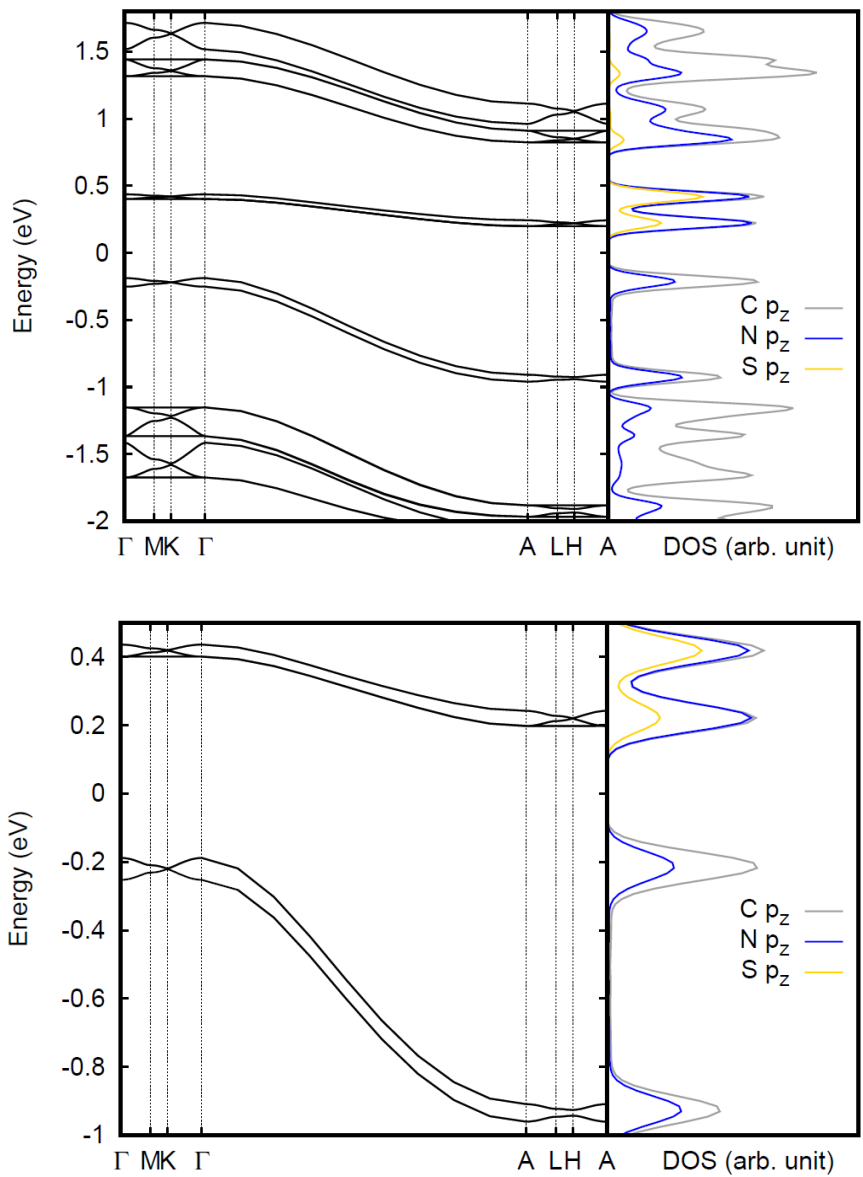

**Figure 5.** DFT calculated electronic band structure for $k_z = 0$ plane and the orbital-resolved projected band. And the projected density of states (PDOS) with blue, grey and yellow lines corresponding to nitrogen, carbon and sulfur element respectively. (Upper) The full-profile and (bottom) the magnified band structures of VBM and CBM.

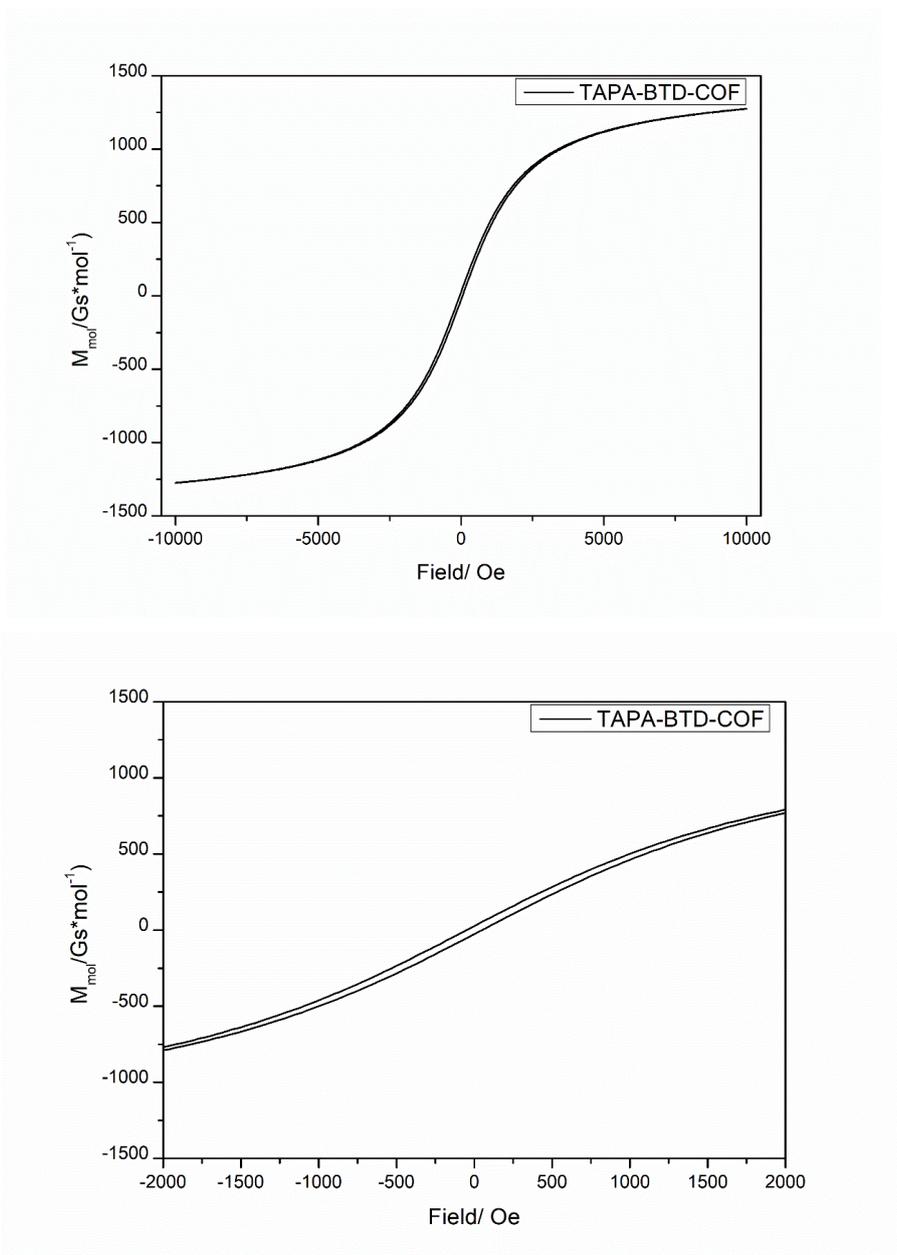

**Figure 6**. Hysteresis loop in M-H curves of TAPA-BTD-COF nanocrystal measured by SQUID MPMS-3 at 300 K in aerobic atmospheric conditions. (Upper) the full-profile M-H spectrum (bottom) the magnified spectrum at low external field.

In our system, because of the deficiency of electron-poor BTD unit, the whole COF structure is electron excessive, which was proven by the electron-paramagnetic-resonance (EPR) characterization in room temperature (300 K) under atmospheric aerobic condition (see Figure 4). There appears a strong signal of electron with intensity about 2000 in EPR spectrum. To investigate its magnetic property, we conducted the M-H magnetization experiments in 300 K under aerobic condition. To our surprise, the as-synthesized TAPA-BTD-COF exhibits pronounced magnetic hysteresis loop during M-H measurement (See Figure 6). To further confirm this result, we conducted parallel measurements of different batches of our TAPA-BTD-COF materials synthesized and stored in different time. To our delight, the M-H hysteresis loop remained almost unchanged despite the choice of different batches COFs materials. To further explore the intrinsic magnetic property of this COF material, the M-T measurements under zero-field cooling and field cooling conditions were undertaken (see Figure 7). The M-T curves were transformed to $\chi_{mol} \sim T$ curve according to the density and molar mass of TAPA-BTD-COF. As shown in Figure 7, the $\chi_{mol}$ of TAPA-BTD-COF exhibits a very large value about 0.028 in 300 K. This value is considerably larger than common paramagnetic material with $\chi_{mol}$ about $10^{-3}$ to $10^{-5}$ level. With the continuously cooling, the $\chi_{mol}$ slowly reduces from 0.028 to 0.025. However, the curve demonstrates a steep jump upwards at about 76.4 K from 0.025 to 0.030. The reason of this upwards jumping is not totally clarified. After the jumping to $\chi_{mol} = 0.030$ at 76.4 K, the COF exhibits the same trend as before the jumping. The $\chi_{mol} \sim T$ slope did not show apparent change. And the $\chi_{mol}$ remains near 0.030 from 76.7 K until cooling down to approximately 27.4 K. Further cooling from 27.4 K to 4.0 K, the $\chi_{mol} \sim T$ slope value continues to increase from 0.030 to 0.037. And the $\chi_{mol}$ value increased almost vertically from 8 K to 4.25 K, increasing from 0.033 to 0.037. This large $\chi_{mol}$ is about the level of ferrimagnetic materials, which belong to strong magnetic materials similar as ferromagnetics with spontaneous magnetization and coercive force but magnetic domain antiparallel aligned with different magnetic moment, which possesses $\chi_{mol}$ in the range of $10^0$ to $10^2$. And the $1/\chi_{mol} \sim T$ curve was also plotted to further analyze its magnetic property. From the $1/\chi_{mol} \sim T$ curve, we did not observe the $T_N$ (Néel temperature), which is critical to assign an antiferromagnetic material. Moreover, from the continuously increasing $1/\chi_{mol} \sim T$ slope value, we can also assign it as a ferrimagnetic material other than antiferromagnetics. The relatively lower $\chi_{mol}$ and coercive force differs it with ferromagnetic materials. From the extrapolating of $1/\chi_{mol} \sim T$ curve to cross with x axis (y axis value equals zero), we obtained the critical temperature $\Delta$ equaling to -33.03 K. This negative value also differentiates it with typical ferromagnetic materials. Due to these phenomena, we consider that our TAPA-BTD-COF belongs to strong magnetic ferrimagnetic material. This is the first stable pure organic non-metal material with considerable M-H hysteresis loop, coercive force and giant $\chi_{mol}$ up to 0.028 (previously aerobic non-stable organometallic magnet mainly in the range of $10^{-6}$ to $10^{-4}$) at room temperature in aerobic condition, and possess a larger $\chi_{mol}$ up to 0.037 at 4.0 K.

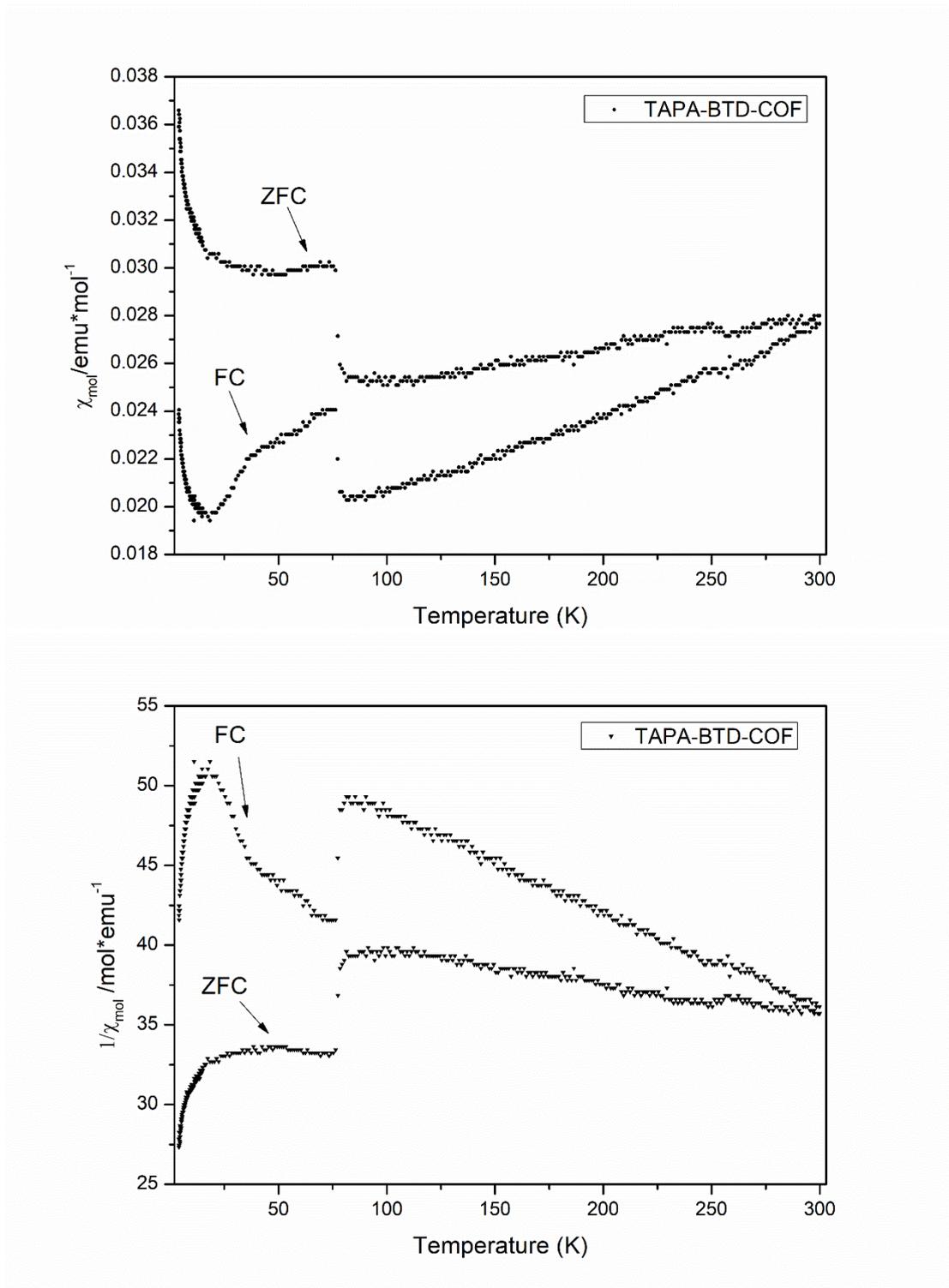

**Figure 7.** (Upper) Zero-Field cooled (ZFC) and field cooled (FC) magnetization curves $\chi_{mol} \sim T$ of TAPA-BTD-COF nanocrystals measured at 200 Oe field. (Bottom) Zero-

Field cooled (ZFC) and field cooled (FC) curves of $1/\chi_{mol}$ ~T of TAPA-BTD-COF nanocrystals measured at 200 Oe field.

## IV. Conclusion

In summary, we reported the discovery of the first non-metal organic ferrimagnetic material TAPA-BTD-COF, which exhibited unconventional strong ferrimagnetism and even keep its strong and stable magneticity in 300 K and aerobic atmosphere with considerable M-H hysteresis, coercive force and $\chi_{mol}$ up to 0.028. The synthesis, characterization and magnetic performances were provided in detail. We discussed the origin of this non-trivial physics phenomena and attributed it to the population of excess electrons locating and localized in CBM flat-band. The electron-populated flat band hosted the location for strong electron correlation. The strongly correlated electron systems lead to this unprecedented ferrimagneticity for our non-metal and non-free-radical organic polymeric material. The further experimental studies and calculations are still under researching in our laboratory. This is the first room-temperature strong magnetic organic materials exhibiting ferrimagnetism with $\chi_{mol}$ up to 0.028 at 300 K in aerobic atmospheric condition.

**Funding:** This work was financially supported by the National Natural Science Foundation of China (Grant Numbers 22076007, 21703005).